\documentstyle[12pt,psfig]{article}
\begin{document}
\begin{center}
{\Large \bf Time Delay Plots of Unflavoured Baryons}
\vskip0.5cm
N. G. Kelkar$^{1,3}$, 
M. Nowakowski$^{2,3}$, K. P. Khemchandani$^1$ and S. R. Jain$^1$\\
{$^1$\it Nuclear Physics Division, Bhabha Atomic Research Centre, \\
Mumbai 400 085, India}
\\
{$^2$ \it Universit\"at Dortmund, Institut f\"ur Physik, D-44221 Dortmund, 
Germany}\\
{$^3$ \it Departamento de Fisica, Universidad de los Andes, 
Cra. 1E No.18A-10, Santafe de Bogota, Colombia} 
\end{center}
\begin{abstract}
We explore the usefulness of the existing relations between 
the $S$-matrix
and time delay in characterizing baryon resonances in 
pion-nucleon scattering. 
We draw attention to the fact that 
the existence of a positive maximum in time delay
is a necessary
criterion for the existence of a resonance 
and should be used as a constraint 
in conventional analyses which locate resonances 
from poles of the $S$-matrix and Argand diagrams. 
The usefulness of the time delay plots of resonances 
is demonstrated through a detailed analysis of the time delay
in several partial waves of $\pi N$ elastic scattering. 
\end{abstract}
\noindent
PACS numbers: 14.20.Gk, 13.85.Dz, 03.65.Nk
\section{Introduction}
Ever since the discovery of the first excited nucleon state 
\cite{fermi}, the baryon resonances have played a major role 
in particle and nuclear physics and have contributed crucially to
the search of the fundamental building blocks of 
nature. We perceive them now as the low energy manifestation of 
quantum chromodynamics (QCD) with three quark degrees
of freedom. 
However, low energy QCD is still not well 
understood and very often one is left with models. The status of the
resonances can differ from case to case as can their parameters 
extracted from different experiments. 
The $N^*$ programs at Jefferson Lab \cite{jefferson} and the
forthcoming Japan Hadron Facility 
\cite{japan} 
have both revived the area. 
%\cite{QCDmodels, QCDmodels2}
The various theoretical studies [4-7] 
%\cite{oset,weise} 
and the hope to find exotic hadrons \cite{exotics}, 
make the area once again an exciting field of physics. It is then justified to 
look at the baryon resonances from
a yet different perspective and to analyze the existing data in a novel albeit
well established way, while waiting for new experimental results.
Specifically, we refer to the time delay method which was introduced
into scattering theory and especially resonance physics 
by Eisenbud and Wigner \cite{eisen,wigner,wigner2}. In view of the considerable 
number of standard textbooks [12-17]
%\cite{brans}-\cite{bzp} 
and numerous papers [9-11,18-39]  
%\cite{eisen, wigner, wigner2}, \cite{smith}-\cite{neels3} 
published since the
seminal paper by Wigner \cite{wigner}, we provide only a short introduction. 
Time delay is a measure of the collision time 
in a scattering reaction which can be calculated 
directly from the phase shift or 
the $T$ matrix. Obviously, such a concept has a 
close connection to the appearance of an unstable intermediate state 
(resonance), which, due to its finite lifetime, ``delays'' the reaction.
Though the interest in time delay ever since the first papers was
unabated, it is only
recently that it has been used in practice in quantum scattering
theory (in chaotic scattering \cite{chaos}, hadron resonances [33,37-39]
%\cite{fraxedas},\cite{neels}-\cite{neels3} 
and heavy ion collisions \cite{hic}) and 
tunneling phenomena \cite{calderon}, with success. The present work is simply
a logical extension of this program carried over to baryon resonances, partly 
done already in \cite{neels}. 
   
To identify the baryon resonances, one performs a partial wave analysis 
of the meson baryon scattering data and obtains the energy dependent 
amplitude (or $T$-matrix) by fitting cross section data. 
Resonances are then determined
by locating the poles of the $T$-matrix on the unphysical sheet and studying 
the Argand diagrams of the complex $T$-matrix. 
Due to model dependence in the analyses of the 
energy-dependent amplitudes, there are differences
in the resonance parameters quoted by different groups 
\cite{pdg}. The resonance receiving confirmation from several analyses
is considered to be well established. 
Though we do not dispute the usefulness of the pole of the $S$-matrix, 
we note that there exist several views in literature, regarding
the definition of a resonance. 
In a review article \cite{dalitz}, Dalitz discussed
various criteria for the existence of a resonance elaborately, with
the conclusion that for the case of a pole in the $S$-matrix, $S(E)$, in the
unphysical E-plane lying sufficiently close to the physical E axis, there
is no ambiguity in the conclusion of the existence of a resonance. 
However, the authors in \cite{calucci} constructed
examples in such a way that a sharp resonance was produced without an
accompanying pole in the unphysical sheet. They noted that even the
inverse correspondence, namely, (pole of the $S$-matrix on the unphysical
sheet) $\rightarrow$ (unstable particle) may be questioned. 
In \cite{amjp}, it was pointed out that  
a peak in the cross section cannot be conclusive evidence of a resonance. 
In \cite{newton}, in addition to time delay, 
the exponential decay law was required as a signal of a genuine resonance 
(this may be in view of the existence of double poles, which would lead 
to a non-exponential decay \cite{bell}).
Cautious remarks on the use of Argand diagrams can be found in 
\cite{lipshutz,argand}.
The many different opinions reflect only the fact that the issue is 
not yet satisfactorily settled.  
Indeed, unstable particles remained to be problematic even until now
\cite{unstable}. 
We make use of the requirement stated in literature
and text books [9-39], namely,  
%\cite{eisen}-\cite{neels3}, 
the formation
of a resonance should introduce a large positive time delay in the
scattering of particles. 
We try to extract resonance parameters from the energy distribution 
of time delay by locating the position of the local maximum and reading off 
the width as advocated e.g. in \cite{widthread}. 
Though the non-resonant background can deform the positive resonant 
structure in the vicinity of a resonance, we do expect some positive 
region around the resonance point, with perhaps a less dominant peak.
This is confirmed by our study.

Starting with the definition of time delay
in terms of the $S$-matrix, we obtain its relation with the $T$-matrix
and scattering phase shifts. We shall first demonstrate
the usefulness of the method with examples of well-known 
$N$ and $\Delta$ resonances. 
Later on we proceed to the analysis of time delay in various 
partial waves of $\pi N$ elastic scattering, using the available
single energy values as well as some energy-dependent forms of
the $T$-matrix.  
Before we move on to the discussion of time delay,
it is important to note that the {\it time delay plots} of the
present work are {\it not} the same as {\it speed plots} 
\cite{hoehl} which have been sometimes referred to as time delay 
plots in literature. 
Speed plots are positive definite by definition. Time delay plots can
also assume negative values and only a positive peak signals a resonance.
In the elastic region, the speed is equal to time delay up to a 
constant factor, but once the inelastic channels open up, this
is no longer true \cite{neels}.

Considering the fact that the time delay method has 
so far not been applied to baryon resonances (but has been
successfully applied to meson resonances \cite{neels3}), our study is 
a practical test of time delay when applied directly to data.
When applied to theoretical $T$ matrix solutions, we could 
say that indeed the model is being 
tested, if we consider the resonances to be well established.

In passing, we note that time delay is also related to the so called
arrival time in quantum mechanics \cite{arrival} and has
also been used to obtain the density of resonances \cite{beth}. 

\section{Time delay in resonant scattering}
We shall now discuss the expressions which quantify time delay and
can hence be used to characterize resonances.
\subsection{Relation to phase shifts}
In the early fifties, using a wave packet analysis, 
Bohm \cite{bohm}, Eisenbud \cite{eisen}
and Wigner \cite{wigner}, obtained an expression for the time delay 
$\Delta t$ in binary collisions. In the case of elastic scattering, 
they derived $\Delta t$ in terms of the energy derivative of the scattering
phase shift as follows:
\begin{equation}\label{1}
\Delta t = 2 \hbar {d\delta \over dE}\,.
\end{equation}
The formation of a resonance in a scattering process, 
introduces a positive time delay between the
arrival of the incident wave packet and its departure from the 
collision region. From the above relation, one expects the phase shift
to increase rapidly in the vicinity of a resonance.

The wave packet analysis of time delay was extended by Eisenbud to 
inelastic collisions \cite{eisen}. He defined the
delay time matrix {\bf $\Delta t$}, such that an element 
$\Delta t_{ij}$ of this matrix, corresponded to the time interval
between the outgoing wave in channel $j$ and the ingoing wave 
in channel $i$. This time delay, $\Delta t_{ij}$, is related to the
$S$-matrix as follows: 
\begin{equation}\label{3}
\Delta t_{ij} = Re \big [ -i \hbar (S_{ij})^{-1} {dS_{ij} \over dE}
\big ] \, .
\end{equation}

Before we proceed further, we note that the phase shifts, in principle,
depend on 
the orbital angular momenta,  
$l$, $l^{\prime}$, of the initial and final states respectively 
and on the total angular momentum $J$. 
However, we have suppressed this dependence in the expressions whenever
not relevant. 
In the present work, we consider $\pi N$ elastic scattering, 
which is the scattering of a spin zero and spin one half particle. Since
the total spin in the final and initial state is $S = S^{\prime} = 1/2$ 
and conservation of parity gives $l = l^{\prime}$, the total angular
momentum $J$ takes the values $l - 1/2$ and $l + 1/2$. The $S$-matrix
is diagonal in $l$ and its elements are related to phase shifts as 
$S_{ll}^J = exp(2 i \delta^J)$, for the elastic case in the absence of
inelasticities.

We see that in the case of purely elastic
scattering ($j=i$), and using a phase shift formulation for the $S$-matrix 
where $S = e^{2i\delta}$, we get,
\begin{equation}\label{5}
\Delta t_{ii} = 2 \hbar {d\delta \over dE}\, ,
\end{equation}
which is the same as Eq. (\ref{1}). 
These $\Delta t_{ii}$ are related to 
the lifetimes of metastable states or resonances in elastic scattering (see
\cite{smith}).
At high energies, where apart from elastic scattering, the possibility
of scattering into inelastic channels also opens up, the elastic $S$-matrix
element is defined as $S = \eta e^{2i\delta}$, where $\eta$ is the 
inelasticity parameter defined such that $0 < \eta \leq 1$. 
Substituting the modified $S$ (i.e. $S =\eta e^{2i\delta}$) in 
Eq. (\ref{3}), gives, 
\begin{eqnarray}
\Delta t_{ii} &=& Re \biggl [ \,-i\hbar\, \biggl( \,2i 
{d\delta \over dE} + {d\eta \over dE}\, {1\over \eta}\, 
\biggr ) \biggr ] \,\,= \,2\hbar \, {d\delta \over dE}\, .
\end{eqnarray}
The above equation is the same as Eqs (\ref{1}) and (\ref{5}). Thus it 
can be seen that the expression for the time delay, 
$\Delta t_{ii}$, for elastic scattering is the same, irrespective of
the presence of inelastic channels.

It is clear from the above expressions that time delay can also
take negative values resulting from phase shifts which decrease as
a function of energy.  
However, the negative delay times cannot assume arbitrarily large values. 
In the case of elastic scattering (for the case of $l=0$ and $1$) it
was shown by Wigner \cite{wigner}, that the causality condition puts a 
constraint on the lower value of the phase shift derivative (related in 
an obvious way to time delay), which in case of high momenta, i.e. for
large $k$ is given as,
$d\delta_l/dk > - a$.
$a$ can be interpreted as the range of the interaction potential. 
We do observe some regions of 
large negative $\Delta t$ which will be discussed in Section 3.

\subsection{Relation to $T$-matrix}
Instead of using the phase shift formulation of the $S$-matrix, we now 
start by defining the $S$-matrix in terms of the $T$-matrix, i.e., 
\begin{equation}\label{6}
S = 1 + 2 i T \,,
\end{equation}
as is usually done in partial wave analyses of resonances
\cite{manley,arndt2}. 
The matrix $T$ contains the entire information of the resonant and
non-resonant scattering and is complex ($T = T^R + iT^I$).
Substituting from Eq. (\ref{6}) into the expression for time delay
in (\ref{3}), 
the time delay $\Delta t_{ii}$, in
terms of the real and imaginary parts of the amplitude $T$ is given as,  
\begin{equation}\label{tmat2}
S^*_{ii} S_{ii}\, \Delta t_{ii} = 2 \hbar \biggl[ {dT^R_{ii} \over dE}
+ 2 T^R_{ii}
{dT^I_{ii} \over dE} - 2 T^I_{ii} {dT^R_{ii} \over dE} \biggr],
\end{equation}
where $S^*_{ii} S_{ii}$ can be evaluated using Eq. (\ref{6}). 
In the present work, we have evaluated the time delay in $\pi N$
elastic scattering and hence, $i$ corresponds to $\pi N$ in the 
above equation.

Although a simple Breit-Wigner (BW) is not always a good choice to
describe a broad hadronic resonance, it is instructive to see
the results we get for time delay, starting from a BW matrix element. 
If we insert one such commonly used form of the
$T$-matrix \cite{arndt2} in resonance regions, namely, 
\begin{equation}
T = {\Gamma/2 \over E_R \,- \,E \,- \,i\Gamma/2}\,,
\end{equation}
in (6), we obtain, 
\begin{equation}\label{9}
\Delta t(E)_{BW} = {\hbar \Gamma \over (E_R - E)^2 + {1 \over 4} \Gamma^2}
\end{equation}
and the time delay at the resonance energy $E_R$ (within the assumption
that the widths are not energy dependent) is, 
\begin{equation}\label{width}
\Delta t(E_R)_{BW} = {4 \hbar \over \Gamma} \, .
\end{equation}
A simple BW $T$-matrix, however, can be misleading, especially while
discussing time delay. The reason among others is that it lacks certain 
usually expected properties 
(threshold behaviour being one of them). 
We shall come to this point in greater detail in section 4.

Before ending this section, we note the dependence of time delay
on wave packets. 
It is well-known that the survival probability and lifetime 
of an unstable quantum state depend on its preparation.  
Explicit formulae including wave packets can be found for unstable 
neutral kaons in \cite{sachs}. We expect a 
similar dependence to be present in the expressions for 
time delay. Indeed, as given in \cite{nusszwg}, 
\begin{equation} \label{wave1}
\Delta t(E)= 8\pi^2 \hbar \int_0^{\infty} dE^{\prime} \vert 
A(E^{\prime},E)\vert^2 2{d\delta \over dE^{\prime}}\, \,,
\end{equation}
where $A(E^{\prime})$ is the initial wave packet in momentum space. 
If the wave 
packet is sharply centered around an energy $E$, we recover Eq. (1).
In scattering processes where one measures the cross sections and 
distributions, the wave packets are indeed narrow, i.e., the energy spread
$\Delta E \ll \Gamma$   
(see the second and last reference in \cite{bransden} for a discussion of this 
issue). Hence we can use Eqs (1-4) to calculate time delay.

In the next section, we shall evaluate the time delay in several
partial waves of 
$\pi N$ elastic scattering. We have checked that the values of 
time delay, $\Delta t_{ii}$, 
obtained either using the derivative of the real phase shifts
as in Eq. (\ref{5}) or the $T$-matrix as in Eq. (\ref{tmat2}) are 
the same. Since both the methods are equivalent, one can in fact use
fits to the single energy values 
\footnote{
The values of phase shifts in different 
partial waves obtained by fitting the cross section data at the 
available energies are known as single energy (SE) values of phase shifts.
The error bars on these phase shifts naturally depend on the errors in
the measured cross sections. The elastic $T$-matrix element 
is related to the phase shift and
inelasticity parameter, $\eta_l$, as: $T_l = (\eta_l \,e^{2i\delta_l} -1)/2i$. 
Thus, one can also obtain SE values of the $T$-matrix.}
of phase shifts to 
extract resonance parameters.  

\section{Time delay plots of resonances in $\pi N$ elastic
scattering}

We now analyze the existing $\pi N$ scattering
data using time delay plots. To demonstrate the usefulness of the method,
we plot time delay in the energy regions where two well-known 
baryon resonances occur.
In Fig. 1 are shown the real and imaginary parts of the complex
$T$-matrices, the phase shifts and the corresponding time delay 
in the $P_{33}$ and $D_{13}$ partial waves in $\pi N$ scattering,  
evaluated using the $T$-matrices (solid lines) 
which fit the single energy values of $T$ very well. 
The filled circles in Fig. 1 are the single energy values of phase
shifts extracted from the cross section data on $\pi N$ elastic
scattering \cite{arndtpn}. 
The widths of the $P_{33}$ and $D_{13}$ peaks at half
\begin{figure}[h]
\centerline{\vbox{
\psfig{file=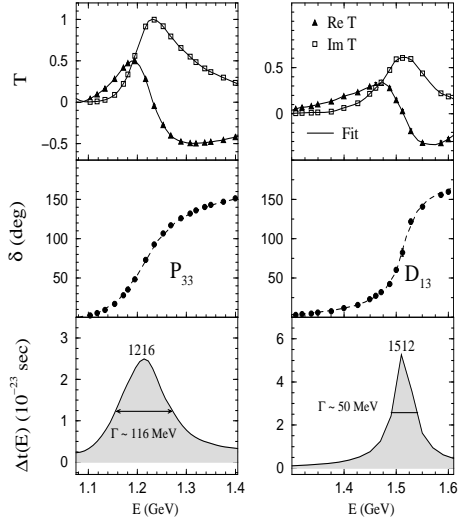,height=7cm,width=6cm}}}
\caption{Single energy values of the real part of the $T$-matrix (filled
triangles), imaginary part of $T$ (open squares), phase shifts (filled circles)
and the time delay $\Delta t$ evaluated in the $P_{33}$ and $D_{13}$
partial waves of $\pi N$ elastic scattering. The time delay is evaluated
using the $T$-matrix given by the solid lines which fit the single
energy values very well.} 
\end{figure}
maximum can be read from Fig. 1 to be around 116 and 50 MeV
respectively. The peaks in the energy distributions occur at
1216 and 1512 MeV respectively. The average values of 
Breit-Wigner masses (widths) given in the Summary
Table (ST) of the Particle Data Group \cite{pdg} for these 
$P_{33}$ and $D_{13}$ resonances are 1232 (120) and 1520 (120) MeV 
respectively. The $\Delta$(1232) decays almost 100\% to the $\pi N$
channel and hence the time delay width seems to be in good agreement
with the above value listed in the ST. The $D_{13}$ has a branching
ratio of 50 to 60\% to the $\pi N$ channel and the width of the
time delay distribution is consistent with the partial width listed
in the ST. 
Thus we see that in the case of purely elastic scattering as well
as in the case of elastic scattering in the presence of inelastic
channels, the method is quite useful. 
The peak position and width of the time delay 
distribution give the
mass and elastic partial width of the resonance, respectively.

Interestingly, the $P_{33}$ phase shift of the only
$\pi N$ resonance ($\Delta (1232)$) in the elastic region, 
remains positive and shows the characteristic resonant jump in
this region. Hence, in this case, the speed 
defined in \cite{hoehl} is the same as time delay up to a constant
factor. 

\subsection{New resonances from single energy values of phase shifts}
We shall
now evaluate time delay from fits to single energy (SE) values of
phase shifts.  
Since the results depend crucially on the quality of the data, 
we chose data sets with small error bars and made separate 
$n^{th}$ order polynomial fits to different energy regions of the phase
shift. It would be more appropriate to consider error
bars and perform a $\chi^2$ fit, with a certain function. However, such a 
procedure would not be able to pick up the small structures and 
would amount to giving results similar to the energy dependent ones. 
We also chose to fit SE values of phase shifts 
rather than the SE values of 
real and imaginary parts of the $T$-matrix, simply as a matter
of convenience. The time delay evaluated using fits to phase shifts
or $T$-matrices is actually the same. The advantage of calculating time
delay from such fits is that the results are directly related to
data. The disadvantage is that they are 
sensitive to the quality of the data and hence to the fit. 
There also exists the well known {\it continuum ambiguity} problem
with the SE values of phase shifts \cite{ambig}. 
However, the present work does not aim at finding solutions to the problems 
related to the extraction of SE values. Hence, we use the 
values as available in literature and check if we still get some
useful results for time delay.

We perform this analysis for the $I = 1/2$ partial 
waves, $P_{11}$, $P_{13}$, $D_{13}$, $S_{11}$ and $F_{15}$ in $\pi N$ elastic
scattering. We note that in spite of the above mentioned 
problems, we get strikingly similar peak positions and widths 
as compared to
the Summary Table resonance parameters.  
\begin{figure}
\centerline{\vbox{
\psfig{file=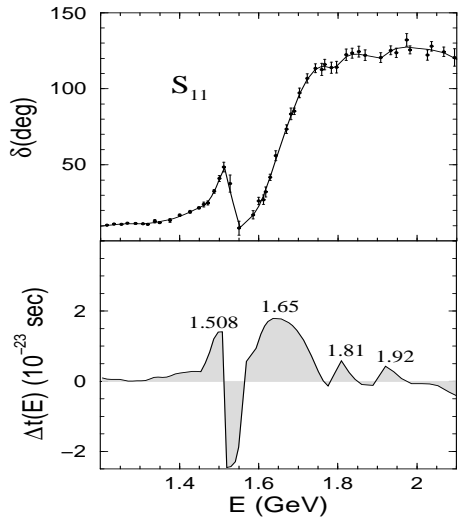,height=7cm,width=6cm}}}
\caption{Single energy values of phase shifts and the corresponding time
delay in the $S_{11}$ partial wave evaluated from a smooth fit
to the phase shifts.}
\end{figure}

The results in Figs 2 and 3 reveal that time delay 
has the following main characteristics: (i) it locates well established
resonances 
(ii) the positive peaks are more prominent
than in the case where we calculated the same quantity for the 
energy-dependent solutions (see section 3.2 below), 
(iii) there exist regions of negative time delay
in addition to the positive peaks 
(iv) the new feature here is that
we find additional resonant peaks. At present, 
given the quality of the data, it is not clear if these new structures are  
artifacts of the fit to the data or genuine indications of (new) resonances.
For example, in the $P_{11}$ and $S_{11}$ cases, 
we have hints for new resonances in the higher energy regions where the
quality of data is worse. 
On the other hand, some one-star 
resonances like $P_{11}(2100)$ and $S_{11}(2090)$ are not excluded. 
We find evidence for the 3-star resonances $F_{15}(2000)$ and 
$P_{13}(1900)$, with some indication 
that the latter consists actually of two nearby resonances.
In the $D_{13}$ partial wave, we observe distinct peaks at 1512, 1695 and 
1940 MeV, which could be associated with the 4-, 3- and 2-star resonances
$N(1520)$, $N(1700)$ and $N(2080)$ respectively. 
Note however that the existence of the two small peaks in this case
depends crucially on two data points, and hence on the fit. These two points 
are sufficiently above continuum to justify the peaks
(more so as they can be associated with known resonances).
There seems to be more structure in the $1400-1700$ MeV region of $S_{11}$.  
A fit made to this detailed structure reveals the possibility of four 
resonances around $1650$ MeV.
Indeed, there is some support
for this structure from recent works in literature \cite{QCDmodels2, Sexp},
where the existence of new resonances at 1.6 and 1.7 GeV is predicted
within quark models. 

With the availability of more precise data on 
cross sections which would enable a better extraction of the
\begin{figure}
\centerline{\vbox{
\psfig{file=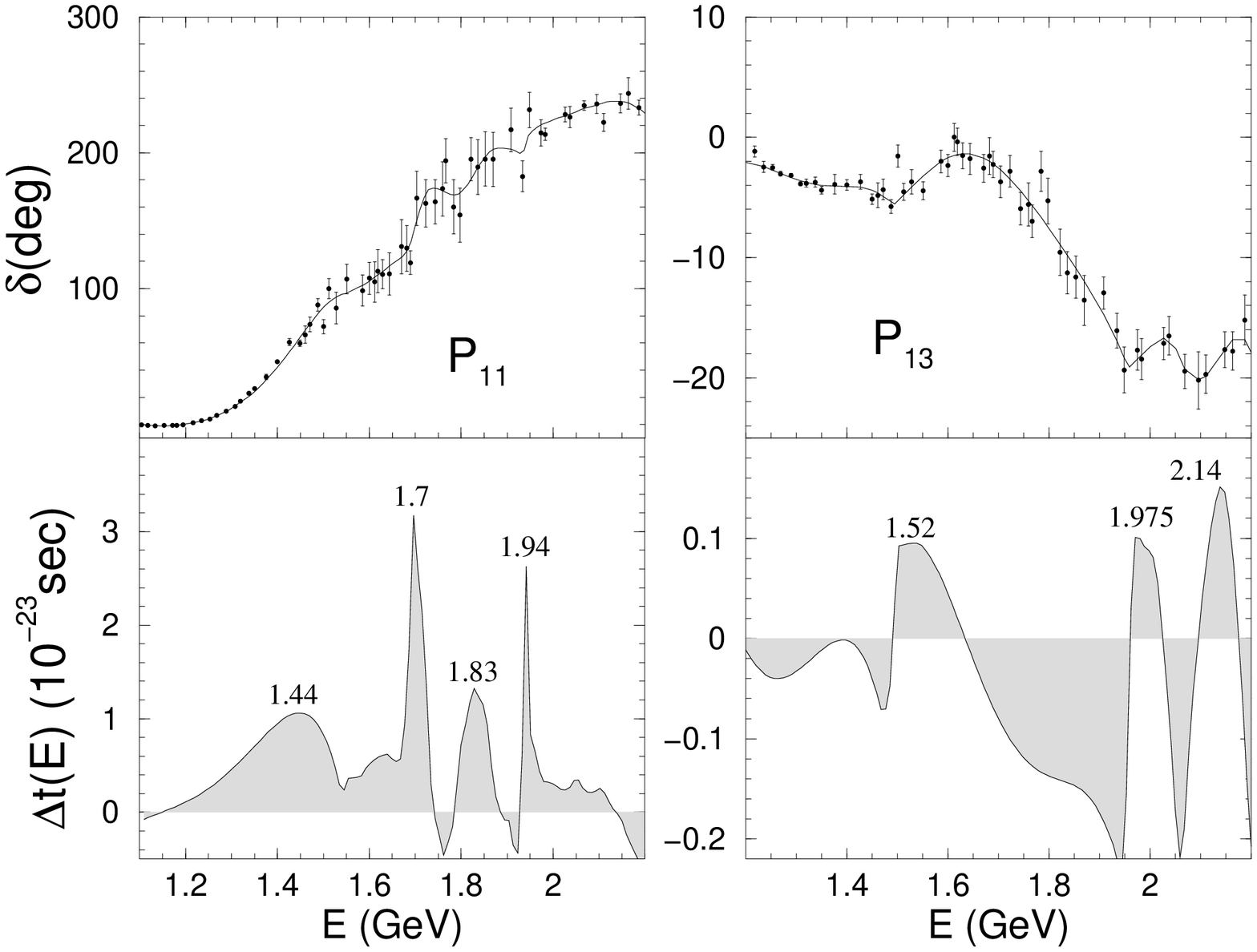,height=8cm,width=8cm}}}
\vskip0.1cm
\centerline{\vbox{
\psfig{file=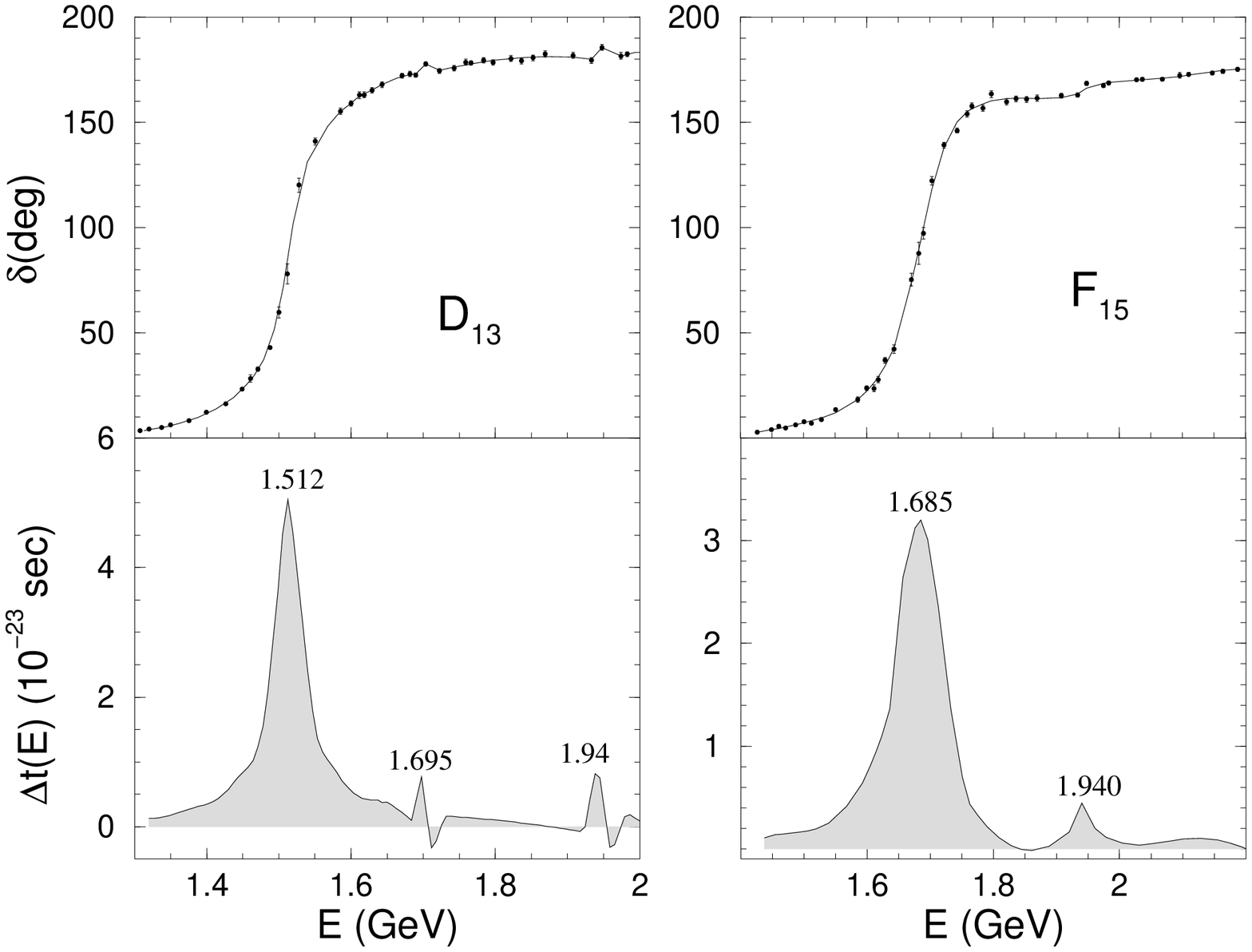,height=8cm,width=8cm}}}
\caption{Single energy values of phase shifts and the corresponding time
delay in the $P_{11}$, $P_{13}$, $D_{13}$ and $F_{15}$ partial waves of
$\pi N$ elastic scattering. The time delay is evaluated using the solid
lines which are fits to the single energy values of phase shifts.} 
\end{figure}
\begin{table}
%\centering
Table 1. Peak positions and corresponding widths from
time delay plots using fits to single energy values of phase
shifts. Masses and widths are in MeV.\\

\begin{tabular}{|c|c|c|c|c|}
\cline{1-5}
           && &&\\
L$_{2I,2J}$&ST average&B.F.=$\Gamma_{\pi N}/\Gamma$ 
&Partial width&Time Delay\\
Mass$^{status}$(Full&Pole (P)&(ST average)&= B.F.$\times$&Peak 
[Partial\\
Width)&Re P [-2 Im P]& &(-2 Im P)&Widths]\\
& & &&\\
\cline{1-5}
D$_{13}$ & &&&\\
1520$^{****}$ [120]&1510 [115] & 0.5 - 0.6& 58 - 69& 1512 [48]  \\
1700$^{***}$ [100]&1680 [100]& 0.05 - 0.15 & 5 - 15&1695 [12]  \\
2080$^{**}$ [-]&1824 - 2120 &    -         & &1940 [18]  \\
           & &&&\\
\cline{1-5}
 F$_{15}$& &&&\\
1680$^{****}$ [130] &1670 [120]& 0.6 - 0.7 &72 - 84 &1685 [91]  \\
2000$^{**}$ [-]& - &    -         & &1940 [33]  \\
           & &&&\\
\cline{1-5}
P$_{11}$ & &&&\\
1440$^{****}$ [350]&1365 [210]& 0.6 - 0.7&126 - 147& 1440 [207] \\
1710$^{***}$ [100]&1720 [230]& 0.1 - 0.2&23 - 46& 1700 [37]  \\
 -	    &     -                &    -         & &1830 [65]  \\
2100$^{*}$ [-]& - &    -         & &1940 [13]  \\
           && &&\\
\cline{1-5}
P$_{13}$ && &&\\
- &          -           &    -         & - & 1520 [101] \\
1900$^{**}$ [-]& &    -        & - & 1975 [51]  \\
     -       &          -           &    -         & - & 2140 [54]  \\  
           & &&&\\
\cline{1-5}
S$_{11}$ && &&\\
1535$^{****}$ [150]&1505 [170]& 0.35 - 0.55&60 - 94 &1510 [37.8] \\
        - &        -             &    -         & &1590 [25.2] \\
1650$^{****}$ [150]&1660 [160]& 0.6 - 0.8&96 - 128&1630 [$\sim$ 40] \\
     - &        -             &    -         && 1680 [$\sim$ 24] \\
     - &        -             &    -         & &1700 [25.6] \\
     - &        -             &    -         & &1810 [31] \\
2090$^{*}$ [-]&1795 - 2220&    -         & &1920 [38] \\
           && &&\\
\cline{1-5}
\end{tabular}
\end{table}
SE values of phase shifts, one could locate the resonances
from time delay plots more accurately. We limit the discussion in this 
section only to the 5 partial waves shown in Figs 2 and 3, since a detailed
analysis with all partial waves would make sense only when the SE 
values of phase shifts would be better known. We list our findings in Table 1.

\subsection{Time delay from energy-dependent amplitudes}
One of the standard methods of characterizing resonances 
involves locating the poles of an energy dependent $T$-matrix
on the unphysical sheet. If these poles correspond to resonances,
a positive maximum in time delay at the energies where the poles occur 
is expected. However, none of the existing analyses
are constrained by this necessary condition. In what follows, we use the 
energy-dependent solutions obtained from the SAID program \cite{arndtpn}  
as an example of such analyses, to evaluate time delay.  
We start with the $I= 1/2$ partial waves in $\pi N$ elastic scattering. 
In Fig. 4a, we plot the SAID solution FA02 of the complex amplitudes.  
The corresponding time delay, using these $T$-matrices and those from 
an earlier analysis (SM95 solutions) by the same group, is plotted in
Fig. 4b. 
The two solutions give rise to similar values of time delay in
all except the $P_{11}$ and $P_{13}$ partial waves, 
where the peak positions differ. The FA02 solutions which are in better
agreement with the SE values as compared to SM95   
were obtained \cite{arndtdiff} using a much bigger database. 
The $P_{11}(~2000)$ peak obtained using SM95 is not seen with FA02. 
Indeed it was noted in \cite{arndtdiff}, that the most significant 
shifts in the pole values occur in the P-waves ($P_{11}$ and $P_{13}$).  
We observe the resonant peaks in the $P_{11}$, $S_{11}$, $D_{13}$, $D_{15}$, 
$F_{15}$ and $P_{13}$ partial waves close to the pole positions 
predicted by the $T$-matrices. However, small peaks in the $G_{17}$ and
$H_{19}$ partial waves  
appear at much lower values than the poles.  
The shifts in the time delay peaks as compared to
the pole values could be due to the presence of a non-resonant background
in these partial waves. Yet another explanation is offered in section 4. 
We give a more detailed discussion of the resonances in various
partial waves now.

\underline{${\rm P}_{11}$-resonances:} The SM95 solution gives a 
broad peak around $1400-1500$ MeV and a prominent peak at $\sim 2050$ MeV 
which could be attributed to the $P_{11}(1440)$ and the less established
$P_{11}(2100)$ respectively. 
With the FA02, the two peaks at $1370$ MeV and
$1745$ MeV are a clear signal of $P_{11}(1440)$ and 
$P_{11}(1710)$ listed in the ST. The FA02 $T$-matrix has two 
closely located poles at 1357 and 1386 MeV. The broad time delay
peak around 1370 could
actually be due to two closely overlapping resonances. The observation 
of the time delay peak at 1370 MeV which is much lower than the
ST value of 1440 MeV is similar to the finding of 
Ref. \cite{H}. In fact, most of the time delay peaks being close
to the pole positions, occur at lower values than the parameters of
the ST. The prominent
peak at 2050 MeV with SM95 is not present in FA02 anymore. 

We also note that $P_{11}$ resonances at 1500 and 1700 MeV were found
in \cite{cutkos} from fits to the energy dependence of the amplitude 
obtained from an older VPI single energy analysis.

\underline{${\rm P}_{13}$-resonances:} 
We see peaks at $1585$ and $1600$ MeV corresponding to the FA02 and
SM95 solutions respectively. There exists a pole of the SM95 solution
at $1717$ MeV which is close to the resonance $P_{13} (1720)$. However
the FA02 pole occurs at a much lower value of $1584$ MeV. 
Though the pole values of the two solutions differ a lot, the time
delay peaks with the two solutions are quite close. 

\underline{${\rm S }_{11}$-resonances:}  
We observe a positive peak around $1494$ MeV, which can be attributed to
$S_{11}(1535)$, again as in the case of 
$P_{11}(1440)$ at a considerably lower value. The positive peak at
$1650$ MeV is a nice manifestation of 
$S_{11}(1650)$. We note another phenomenon which commonly occurs 
in time delay plots now.  
The small peak at 1494 MeV is followed by a large negative region (which, 
as emphasized before, cannot be associated with resonance formation).  
The negative time delay is associated with the opening up of new channels
in $\pi N$ scattering and in fact, the minimum in the dip occurs at
the energy corresponding to maximum inelasticity. 
The energy derivative of the phase shift (and hence time delay) is
related through the Beth-Uhlenbeck formula to the change in density
of states in the presence of interaction \cite{beth}. The negative
time delay corresponds to situations where due to the interaction,
the density of states is less than in the absence of interaction. 
This is exactly what happens when the inelastic channels open up
and there is a loss of flux from the elastic channel, thus reducing
the density of states due to interaction.
A detailed discussion on
this issue can be found in \cite{neels}. A repulsive 
non-resonant background could make an additional contribution to the
negative time delay. 

\begin{figure}
\centerline{\vbox{
\psfig{file=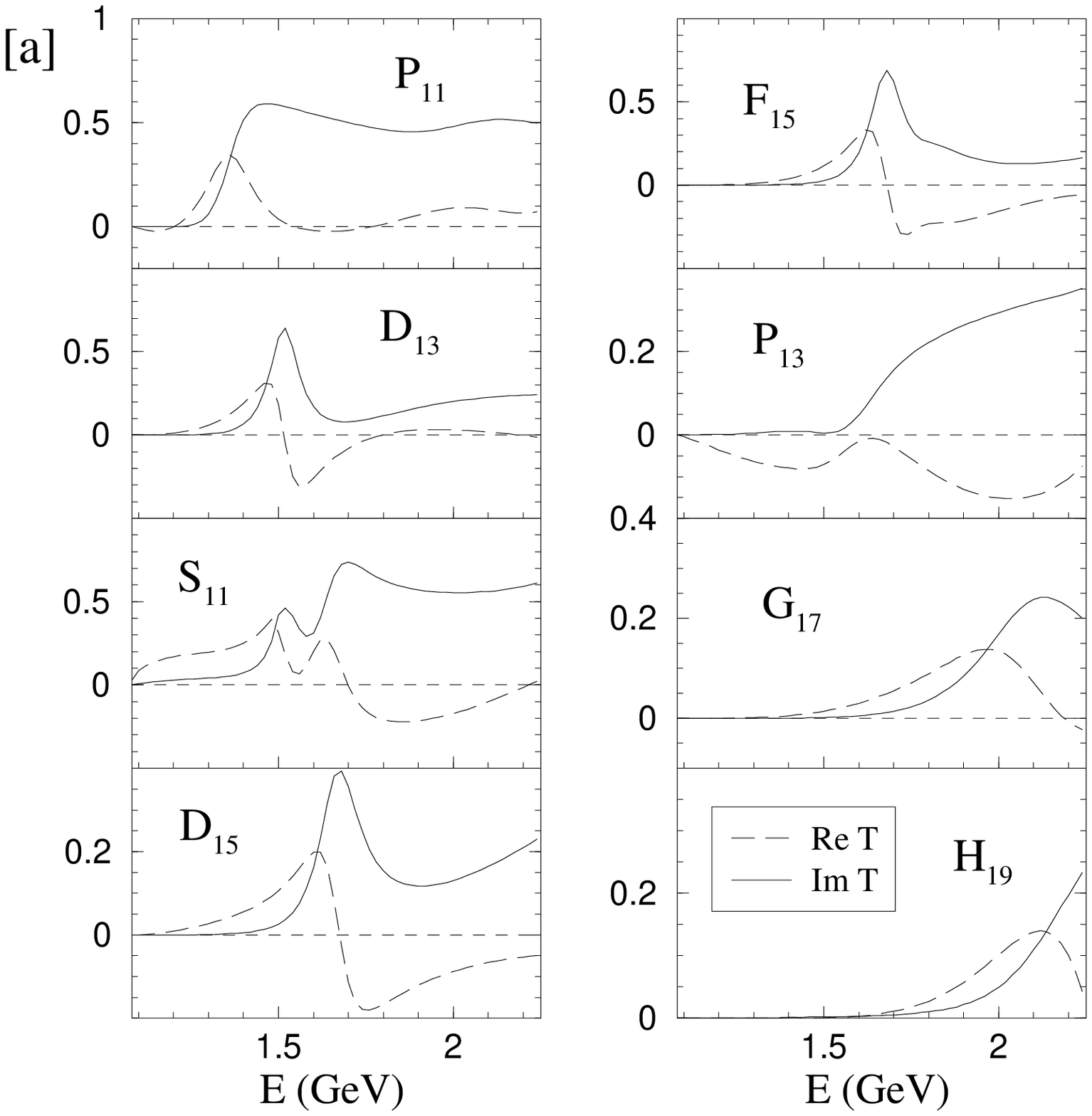,height=8cm,width=8cm}}}
\vskip0.1cm
\centerline{\vbox{
\psfig{file=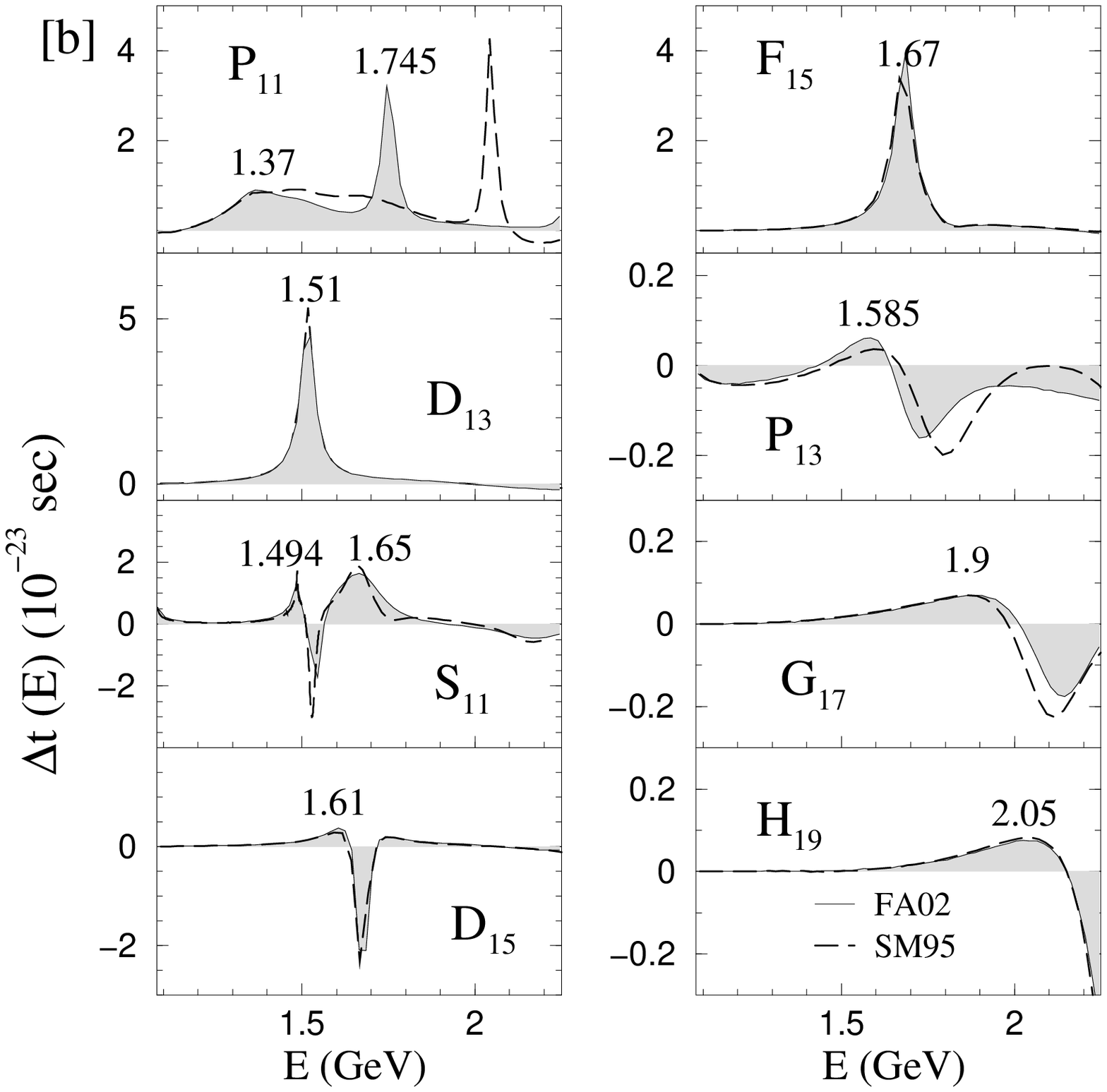,height=8.5cm,width=8cm}}}
\caption{(a) Real (dashed lines) and imaginary (solid lines) parts of the 
$T$-matrix solutions FA02 \protect \cite{arndtpn} 
for isospin, $I = 1/2$, partial waves. 
(b) The time delay (solid lines) evaluated
using the $T$-matrix solutions FA02 (shown in (a)) and an earlier
version SM95 (dashed lines) by the same group.}  
\end{figure}
\begin{table}
%\centering
Table 2. Comparison of nucleon resonance parameters from time delay evaluated
using the FA02 $T$-matrix solution, with the pole positions \cite{fa02} of the
same $T$-matrix, Summary Table values and Speed Plot poles 
(P = E - $i \Gamma /2$) .\\  
\vspace{0.1cm}
\begin{tabular}{|c|c|c|c|c|}
\cline{1-5}
%L$_{2I,2J}$ & Speed plots & FA02 & Branching fraction & Time delay \\
L$_{2I,2J}$ & Speed Plot & FA02 & Branching & Time delay \\
Mass (Full - & Pole(P) & Pole (P) & to $\pi$N   & Peak [Partial  \\
width) & Re P [-2 Im P] & Re P [-2 Im P] &decay mode & - width] \\
\cline{1-5}  
P$_{11}^{****}$ & 1385 [164] & 1357 [162] & 60 - 70 $\%$ & 1370 [298] \\
1440 (350)      &            & 1386 [170] &              &  $\dagger$  \\
      &            &            &              &            \\
\cline{1-5}  
D$_{13}^{****}$ & 1510 [120] & 1514 [103] & 50 - 60 $\%$ & 1510 [50]  \\
1520 (120)      &            &            &              &            \\
\cline{1-5}  
S$_{11}^{****}$ & 1487 [ - ] & 1516 [123] & 35 - 55 $\%$ & 1494 [48]  \\
1535 (150)      &            &            &              &            \\
\cline{1-5}  
S$_{11}^{****}$ & 1670 [163] & 1639 [155] & 55 - 90 $\%$ & 1650 [145] \\
1650 (150)      &            &            &              &            \\
\cline{1-5}  
D$_{15}^{****}$ & 1656 [126] & 1664 [141] & 40 - 50 $\%$ & 1610 [93]  \\
1675 (150)      &            &            &              &            \\
\cline{1-5}  
F$_{15}^{****}$ & 1673 [135] & 1677 [121] & 60 - 70 $\%$ & 1670 [75]  \\
1680 (130)      &            & 1778 [215] &              &            \\
      &            &            &              &            \\
\cline{1-5}  
D$_{13}^{*** }$ & 1700 [120] &       -    &  5 - 15 $\%$ &   -        \\
1700 (100)      &            &            &              &            \\
\cline{1-5}  
P$_{11}^{ ***}$ & 1690 [200] &    -       & 10 - 20 $\%$ & 1745 [50] \\
1710 (100)      &            &            &              &            \\
\cline{1-5}  
P$_{13}^{ ***}$ & 1686 [187] & 1584 [287] & 10 - 20 $\%$ & 1585 [124] \\
1720 (150)      &            &            &              &            \\
\cline{1-5}  
P$_{11}^{ ** }$ &    - & 2009 [458] & -& -  \\
2100 ( - )      &&&&\\
\cline{1-5}  
G$_{17}^{****}$ & 2042 [482] & 2084 [453] & 10 - 20 $\%$ & 1900 [310] \\
2190 (450)      &            &            &              &            \\
\cline{1-5}  
H$_{19}^{****}$ & 2135 [400]    & 2230 [553]  & 10 - 20 $\%$ & 2050 [276] \\
2220 (400)      &               &             &              &            \\
\cline{1-5} 
\end{tabular}\\
\\
$\dagger$ Width is large due to the possibility of 2 overlapping resonances.\\
\end{table}

\underline{${\rm D}_{13}$ and ${\rm F}_{15}$-resonances:} 
Time delay reveals  
the $D_{13} (1520)$ and ${\rm F}_{15}(1680)$ resonances. 
The three- and two-star resonances
$D_{13} (1700)$ and $F_{15} (2000)$ respectively, do not appear in
the time delay evaluated with the SM95 or FA02 solutions. This is
consistent with the fact that the same solutions do not locate
the corresponding poles. However, these
two resonances do appear in the time delay plots obtained from fits
to the single energy values of the $T$-matrix.

\underline{${\rm D}_{15}$ resonance:} 
Though the peak is not prominent, its position and width are close
to the pole values. 

\underline{${\rm G}_{17}$ and ${\rm H}_{19}$-resonances:}
The peak positions in time delay are at much lower values as 
compared to the poles and much broader than the partial widths
corresponding to the poles. 
It is not clear if such a large shift in these 2 cases should be
attributed to the non-resonant background, since most of the other
resonances were shifted from the pole values by only few tens
of MeV at the most. 
Although a remote possibility, it could be that the shifted peaks 
represent resonances
without corresponding poles and the poles at higher values do not
correspond to resonances. These cases could be realistic examples
similar to the constructed ones in \cite{calucci}, where the authors
doubt the one-to-one correspondence between an unstable state and
$S$-matrix pole. A good knowledge of the non-resonant background would 
clarify the situation.  
An alternative explanation will be given in section 4.
\begin{figure}
\centerline{\vbox{
\psfig{file=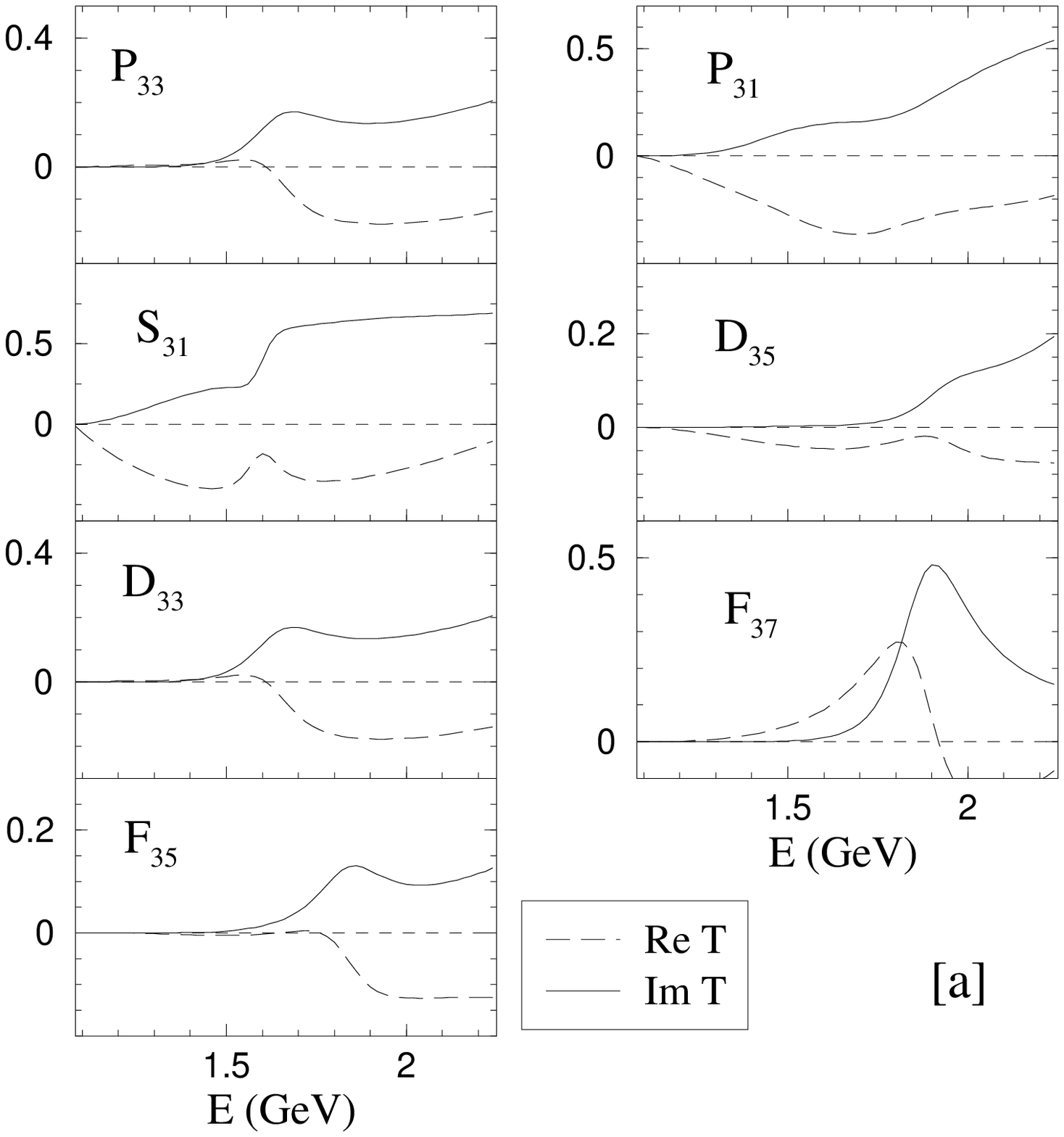,height=8cm,width=8cm}}}
\vskip0.1cm
\centerline{\vbox{
\psfig{file=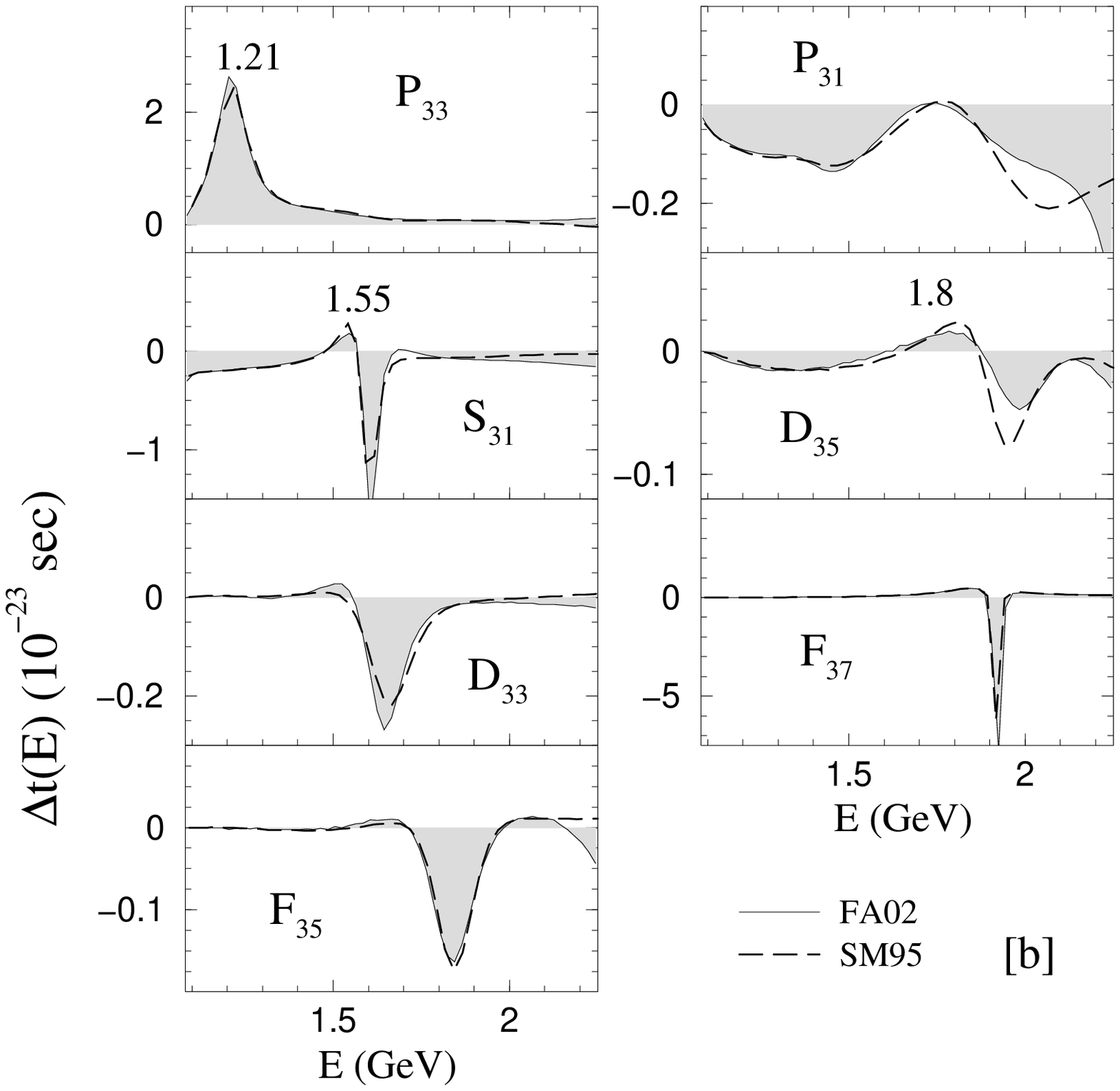,height=8.5cm,width=8cm}}}
\caption{(a) Real (dashed lines) and imaginary (solid lines) parts of the 
$T$-matrix solutions FA02 \protect \cite{arndtpn} for isospin, 
$I = 3/2$, partial waves. 
(b) The time delay evaluated
using the $T$-matrix solutions FA02 (shown in (a)) and an earlier
version SM95 by the same group.}  
\end{figure}
\begin{table}
%\centering
Table 3. Comparison of $\Delta$ resonance parameters from time delay evaluated
using the FA02 $T$-matrix solution, with the pole positions \cite{fa02} of the
same $T$-matrix, Summary Table values and Speed Plot poles 
(P = E - $i\Gamma /2$).\\  
\vspace{0.1cm}
\begin{tabular}{|c|c|c|c|c|}
\cline{1-5}
L$_{2I,2J}$ & Speed plots & FA02 & Branching & Time delay \\
Mass (Full - & Pole (P) & Pole (P)& to $\pi$N   & Peak [Partial \\
width)  & Re P [-2 Im P] & Re P [-2 Im P] &decay mode &- width] \\
\cline{1-5}
%\cline{1-6} 
P$_{33}^{****}$   & 1209 [100] & 1211 [101] & $\geq$ 99 $\%$ & 1210 [108] \\
1232 (120)        &            &            &                &            \\
\cline{1-5}
P$_{33}^{ ***}$   & 1550 [ - ] &    -       & 10 - 25 $\%$   &     -      \\
1600 (350)        &            &            &                &            \\
\cline{1-5}
S$_{31}^{****}$   & 1608 [116] & 1594 [114] & 20 - 30 $\%$   & 1550 [54]  \\
1620 (150)        &            &            &                &            \\
\cline{1-5}
D$_{33}^{****}$   & 1651 [159] & 1633 [254] & 10 - 20 $\%$   &  $\dagger$ \\
1700 (300)        &            &            &                &            \\
\cline{1-5}
S$_{31}^{*** }$   & 1780 [ - ] &     -      & 10 - 30 $\%$   &      -     \\
1900 (200)        &            &            &                &            \\
\cline{1-5}
F$_{35}^{****}$   & 1829 [303] & 1832 [239] &  5 - 15 $\%$   &$\dagger$    \\
1905 (350)        &            &            &                &            \\
\cline{1-5}
P$_{31}^{****}$   & 1874 [283] & 1781 [493] & 15 - 30 $\%$   &      -     \\
1910 (250)        &            &            &                &            \\
\cline{1-5}
P$_{33}^{ ***}$   & 1900 [ - ] &      -     &    -           &      -     \\
1920 (200)        &            &            &                &            \\
\cline{1-5}
D$_{35}^{****}$   & 1850 [180] & 1918 [277] & 10 - 20 $\%$   & 1800 [157] \\
1930 (350)        &            &            &                &            \\
\cline{1-5}
F$_{37}^{****}$   & 1878 [230] & 1871 [236] & 35 - 40 $\%$   &$\dagger$  \\
1950 (300)        &            &            &                &            \\
\cline{1-5}
\end{tabular}\\
\\
$\dagger$ Resonance signal not very clear 
\end{table}

A comparison of the time delay peaks (evaluated using the FA02 solution)
with the ST values and pole positions of FA02 and 
Speed Plots is given in Table 2. The branching fractions of the various
resonances to the $\pi N$ decay channel are also listed. The widths 
in time delay are the partial widths corresponding to the $\pi N$ decay
mode. Widths listed in the second and third columns are full widths.  
It can be seen that we get distinct resonance signals even for resonances
with a branching fraction as small as 10-20 \% to the $\pi N$ channel.

Next, we move on to the analysis of the $I = 3/2$ partial waves in
$\pi N$ scattering (Fig. 5). 
Many conventional pole positions of the energy-dependent $T$-matrices 
occur in regions of negative time delay. However, the time delay plots made
from fits to the single energy values of the amplitude 
do show small peaks due to the $\Delta$ resonances with a small branching
fraction to the $\pi N$ decay mode \cite{neels}. With the exception of
$P_{31}$ where we do not find any positive region, 
the overall picture is similar to the case of the $I=1/2$ resonances. 
It is however fair to say that in the cases 
of $D_{33}$, $F_{35}$ and $F_{37}$ partial waves, the positive peaks 
are really too tiny to be conclusive.
A comparison of the time delay peaks in the $I = 3/2$ partial
waves (evaluated using the FA02 solution)
with the Summary Table values and pole positions of FA02 and
Speed Plots is given in Table 3.

\section{Time delay and Breit-Wigner amplitudes}
In 2.2, we evaluated the time delay from a simple
Breit-Wigner (BW) amplitude, assuming that the branching ratio for 
the elastic channel under consideration is $1$. 
We discuss this topic at this point again as we wish to interpret 
the results of our time delay analysis using $T$-matrix solutions which 
involve BW-like functions.

Generalizing the case of the BW amplitude by considering
an elastic channel $i$ with a branching ratio, 
$Br \equiv \Gamma_i / \Gamma$, 
smaller than one, we have 
\begin{equation} \label{new2}
T={\Gamma_i/2 \over E_R -E -i\Gamma/2}
\end{equation}
from which, after taking into account that the $S$-matrix is $S=1+2iT=\eta 
e^{2i\delta}$, the phase shift $\delta$ can be calculated to be
\begin{equation} \label{new3}
\delta = {1 \over 2}\tan^{-1}\biggl [{\Gamma_i(E_R-E) \over
(E_R -E)^2 +\Gamma^2/4 -\Gamma_i \Gamma/2}\biggr ]\, .
\end{equation}
The derivative of the phase shift taken at $E=E_R$ is, 
\begin{equation} \label{new4}
{d \delta \over dE}\biggr|_{E=E_R}= {1 \over \Gamma}
{Br \over (Br -1/2)}\, .
\end{equation}
From Eq. \ref{new4}, we see that if $Br < 1/2$, time delay at
resonance is negative. The negative region around 
$E_R$ is a local negative minimum
accompanied on two sides by local positive maxima. 
We refer to this phenomenon as `two-horn' structure. 
We can either say that the time delay concept loses its meaning in resonance
physics in channels with a branching ratio less than $1/2$, or, the simple
Breit-Wigner is not an adequate description for a resonant amplitude; although
for many purposes other than time delay, a reasonable
approximation. Indeed, Eq. (\ref{new2}) lacks the threshold factor and 
energy dependent widths. 
The threshold should at least be implemented
correctly by $\Gamma_i \to \Gamma_i [q(E)/q(E_R)]^
{2l+1}$ (see e.g. \cite{meold} and references therein), 
where $q$ is the momentum of one of the initial particles in the CM frame.
This $\Gamma_i$ is not unique and can be modified \cite{arndt2}.
Moreover, as noted in \cite{arndt2} the rest of the energy dependence can
take many different forms (see \cite{zheng} for the latest collection of
such resonant amplitudes). Hence, any argument in connection with time delay
relating the latter with a Breit-Wigner should not be based on (\ref{new2}). 
This is exactly what we have
done by computing the time delay from $T$-matrix solutions which are based
on a more sophisticated BW form \cite{arndt2}. 
Certainly in some cases, an almost `two-horn' structure is visible
and is reminiscent of a BW. 
However, not always should this fact be considered
as a drawback for the time delay concept. 
Indeed, in spite of an almost
`two-horn' structure for $P_{13}$, we find a mass of $1585$ MeV 
as compared to the pole value of $1584$ MeV. 
The $S_{11}$ partial wave, in spite of the almost 
`two-horn' structure, displays two bona fide resonances at their expected 
values. This shows that a more sophisticated BW 
can indeed account for resonances in the time delay.  
The $D_{15}$, $S_{31}$ and $D_{35}$ resonances are in agreement with the 
values found in \cite{arndt2}. 
We do find the $P_{11}(1710)$ in spite of a branching ratio of $10-20\%$. 
Hence, this again demonstrates that arguments based on (\ref{new2}) are
not at all stringent.

There are certainly problematic cases such as the 
$D_{33}$, $F_{37}$ and $F_{35}$.
Dismissing the concept of time delay on account of these cases would 
amount to the same as dismissing the solutions of \cite{arndt2} which also 
fail to find several important resonances quoted in PDG (in 
addition to the fact that the resonance parameters do not always agree with 
the mean PDG value).
This is clearly unacceptable as it is rather the rule 
than exception that different analyses in hadronic resonances 
yield different results.
The fact that time delay is indecisive in certain
cases could be due to the BW used in the parametrization of the 
$T$-matrix solutions. Although much better than
(\ref{new2}) (as already proved by the time delay method itself), it 
might still not be the most general and suitable form for broad
resonances \cite{BWW}.
In the $i \to j$ channel, the numerator of the BW amplitude gets 
replaced by $c_i c_j$ with $c_i=\sqrt{\Gamma_i/\Gamma}$. It was noted
in \cite{kaminski} in connection with $\pi \pi$ scattering, 
that in order to obtain the correct $S$-matrix, the $c_i$'s
should be complex. 
This should apply equally well to the baryon
resonances (a similar argument can be found in \cite{brans}). 
It is worth noting that in our time delay analysis of 
meson-meson scattering \cite{neels3}, in all cases 
with $Br \ll 1/2$ for the elastic channel we found 
positive peaks (and no `two-horn' structures). These cases are not isolated 
as there are six of them. Hence in contrast to the simple theoretical example
of an oversimplified BW, in reality (at least for the meson-meson case) time 
delay works. 
We therefore suspect that 
an improvement on the BW in the baryon case will also lead to an improvement 
of the corresponding time delay results.

Finally, we also examine the influence of the non-resonant
background using the BW parametrization for coupled channels
as in \cite{badalian}. Assuming a diagonal 
background, the $S$-matrix for elastic scattering ($i=j$) is given as,
\begin{equation}
S = e^{2i\eta_i} \biggl [ 1 - 2i {E_R \Gamma_i(E) \over E^2 - E_R^2 + 
i E_R \Gamma(E)} \biggr ]
\end{equation}
where $\eta_i$ is the background phase. In this case, 
\begin{equation}
{d\delta \over dE}\biggr|_{E=E_R} = {Br \over \Gamma(E_R) (Br - 1/2)} 
+ {d\eta_i \over dE}\biggr|_{E=E_R}
\end{equation}
which generalizes Eq. (\ref{new4}).
 
\section{Summary}

Though the resonances in $\pi N$ scattering belong to 
one of the oldest topics of particle and nuclear physics, their study
is far from being 
complete. This can be seen alone from their classification into
four-, three-, two-, and one-star resonances according to the status of being 
well or less established. Disagreements in the extractions of resonance 
parameters are often encountered and theoretical model calculations 
sometimes predict new resonances.

Motivated by statements in literature that a positive time delay is a 
necessary condition to confirm a resonance, in the present work  
we have put the time delay method to test 
and have presented a systematic survey of time delay plots for 
the $\pi N$ resonances. 
In case of a clear signal, the resonance parameters could be extracted
from the time delay plots.
When $\Delta t(E)$ was calculated from the $T$-matrix solutions, we did not 
always find resonances at the energies corresponding to the poles
of the $T$-matrix. 
This might be due to the
model dependence of the $T$-matrix solutions or the non-resonant
background in them. The calculation of $\Delta t(E)$ directly from data
via phase shifts also characterized several established resonances. The
results indicate that in some cases a known resonance 
could actually be a mixture of two neighbouring resonances. 
Detailed fits to the structure in the data revealed new resonances. For
example, in the much talked about $S_{11}$ partial wave 
\cite{QCDmodels2,Sexp,s11} we found some new resonances with some of them
in agreement with the predictions in literature \cite{Sexp}.  
We believe that given 
more precise data, the time delay approach to resonances 
would be a very useful tool to characterize resonances.   
\vskip0.5cm
\centerline{\bf APPENDIX}

We briefly discuss two concepts related to time delay. These were
not used in the present work, but we include them 
to avoid confusion among different concepts.

Time delay is closely related to the lifetime matrix {\bf Q}, 
defined by Smith \cite{smith}. This {\bf Q} is related to the
scattering matrix {\bf S} as, 
${\bf Q} = -i\hbar\, {\bf S}\, d{\bf S}^{\dag}/dE \, .$
The average time delay can be defined as a 
weighted average of the delay times $\Delta t_{ij}$ 
and is the same as $Q_{ii}$. Since the particle
has probability $|S_{ij}|^2$ of emerging in the $j^{th}$ channel,
the average time delay for a particle injected in the $i^{th}$ channel
is given as,
\begin{eqnarray}
\langle\, \Delta t_i \,\rangle_{av} &=& \sum_j \, S_{ij}^* S_{ij} \,
\Delta t_{ij} \,\,
= \,Re \biggr[ \,-i\hbar \,\sum_j S^*_{ij} \,{dS_{ij} \over dE}
\biggr] \,\,
= \, Q_{ii}\,. \nonumber
\end{eqnarray}

Strictly speaking, equations (1), (6) and 
(\ref{wave1})
represent a time ``delay" due to interaction. One can also derive an 
expression \cite{nusszwg} (quoted here for $J=0$): 
$$
T(a)=8\pi^2\hbar \int_0^{\infty} dE^{\prime}\vert A(E^{\prime},E)\vert^2 \left 
\{2{d\delta \over dE^{\prime}}
+2a- {\sin[2(\delta +k^{\prime}a)] \over k^{\prime}} \right \} \, ,
$$
which is interpreted as the time spent within the spherical 
interaction region of radius $a$ in the presence of interaction. 
Since the above equation can be rewritten
as an integral over a probability, $T(a)$ is always positive definite in 
contrast to (1), (6) and (\ref{wave1}).

We also note that
yet another discussion of time delay in classical and quantum 
scattering theory can be found in \cite{hn}. 

\vskip0.5cm
{\bf Acknowledgements: }
The authors wish to thank R. A. Arndt for useful discussions and 
I. Strakovsky for providing the pole values of their $T$-matrix 
solutions.

\end{document}